\newcommand{\be}{\begin{eqnarray}}
\newcommand{\ee}{\end{eqnarray}}

\newcommand{\bea}{\begin{eqnarray}}
\newcommand{\eea}{\end{eqnarray}}
\newcommand{\beas}{\begin{eqnarray*}}
\newcommand{\eeas}{\end{eqnarray*}}
\documentclass[twocolumn,aps,showpacs]{revtex4}
\usepackage{graphics}
\usepackage{epsfig}
\usepackage{amsmath}
\usepackage{amssymb}
\usepackage{amsthm}
\begin{document}
\title{Feynman parametrization and Mellin summation at finite temperature}  
\author{Alejandro Ayala$^{1,3}$, Gabriella Piccinelli$^2$, Angel S\'anchez$^3$
and Maria Elena Tejeda-Yeomans$^4$}   
\affiliation{$^1$Instituto de Ciencias Nucleares, Universidad
Nacional Aut\'onoma de M\'exico, Apartado Postal 70-543, M\'exico
Distrito Federal 04510, Mexico.\\
$^2$Centro Tecnol\'ogico, FES Arag\'on, Universidad Nacional Aut\'onoma de
M\'exico, Avenida Rancho Seco S/N, Bosques de Arag\'on, Nezahualc\'oyotl,
Estado de M\'exico 57130, Mexico.\\
$^3$Instituto de F{\'\i}sica y Matem\'aticas,
Universidad Michoacana de San Nicol\'as de Hidalgo, Apartado Postal
2-82, Morelia, Michoac\'an 58040, Mexico.\\
$^4$Departameto de F\1sica, Universidad de Sonora, Boulevard Luis
Encinas J. y Rosales, Colonia Centro, Hermosillo Sonora 83000, Mexico.}

\begin{abstract}

We show that the Mellin summation technique (MST) is a well defined and useful
tool to compute loop integrals at finite temperature in the imaginary-time
formulation of thermal field theory, especially when interested in
the infrared limit of such integrals. The method makes use of the
Feynman parametrization which has been claimed to have problems when
the analytical continuation from discrete to arbitrary complex values of the
Matsubara frequency is performed. We show that without the use of the MST,
such problems are not intrinsic to the Feynman parametrization but instead,
they arise as a result of $(a)$ not implementing the periodicity brought about
by the possible values taken by the discrete Matsubara frequencies before the
analytical continuation is made and $(b)$ to the changing of the original
domain of the Feynman parameter integration, which seemingly simplifies the
expression but in practice introduces a spurious endpoint singularity. Using
the MST, there are no problems related to the implementation of the
periodicity but instead, care has to be taken when the sum of denominators of
the original amplitude vanishes. We apply the method to the computation of
loop integrals appearing when the effects of external weak magnetic fields on
the propagation of scalar particles is considered.\\  

\noindent Preprint Number UMSNH-IFM-F-2008-16.

\end{abstract}


\maketitle

\date{\today}

\section{Introduction}

At finite temperature, unlike in vacuum, momentum dependent loop integrals in
general depend separately on the time ($p_0$) and space (${\bf p}$)
components of the momentum $P^\mu=(p_0,{\bf p})$, since Lorentz invariance is
lost due to the presence of the medium. As a consequence, the limiting
behavior of these integrals, as the momentum components approach given values,
may depend on the way the limit is taken. For instance, in one loop
self-energy calculations, the {\it infrared limit} $(p_0=0,p=|{\bf
p}|\rightarrow 0)$, that accounts for the plasma screening properties, does
not necessarily coincide with the {\it static limit} $(p_0\rightarrow 0
,p=|{\bf p}|=0)$, that accounts for the long wavelength plasma oscillations. In
physical terms, this non-analyticity as $P^\mu\rightarrow 0$ is due to the cut
structure of the self-energy at finite temperature, where branch cuts appear 
representing scattering processes not allowed in vacuum.  

The above behavior was originally not fully recognized since there were
results indicating that the aforementioned two limits commuted when computing
the real part of the self-energy in a $\phi^3$ theory~\cite{Gribosky,
Bedaque}. In the imaginary time formalism (ITF), the problem with these
calculations was traced back to an incorrect analytic continuation when the
discrete frequency takes on arbitrary complex values~\cite{Weldon}. Since the
erroneous result was obtained from an analysis based on the use of the Feynman
parametrization at finite temperature, it is often thought that such
parametrization is also endemic to the source of the error. 

In the ITF, the in\-tro\-duc\-tion of the Feyn\-man parametrization for the
computation of loop integrals containing two propagators can be
avoided since there are techniques that allow to perform the sum over
Matsubara frequencies in a straightforward manner. However, it has been
recently shown that when studying the influence of weak magnetic fields over
physical processes at finite temperature, the loop integrals that appear
involve products of powers of two or more propagator-like
denominators~\cite{nosotros}. Although it is possible to generalize the
standard techniques to carry out the sum over Matsubara frequencies from the
product of two propagators to the case a of product of powers of these, the
calculations become extremely cumbersome. It is therefore desirable to have a
more direct method to perform these calculations. One of such is the
Mellin summation technique (MST). The method~\cite{Bedingham} calls for the
use of Feynman parametrization, which allows to condense products of powers of
propagators into a single propagator-like factor raised to some power. This is
particularly useful when one seeks an answer in terms of a power series
involving a small parameter, for instance, at a high temperature $T$, the
ratio $m/T$, where $m$ is the particle's mass. 

In light of the well known mishaps with the use of the Feynman
parametrization~\cite{Weldon} in finite temperature calculations, it is
important to establish that the MST in the ITF has the correct analytical
properties when the discrete Matsubara frequency is continued to arbitrary
complex values. In this work we undertake such study. We perform an explicit
calculation of the one-loop self-energy in a $\phi^3$ theory as a guiding tool
to find out how the Feynman parametrization should be used in the ITF of
thermal field theory. The work is organized as follows: In Sec.~\ref{II} we
give a brief summary of the previous analyses that have dealt with this
problem and compare their results. In Sec.~\ref{III}, we give a detailed
derivation of the standard calculation using conventional techniques. We
explore the case when the external momentum approaches zero and 
in particular, give the explicit result in the infrared limit. As we want to
compare with the MST, the answer is given in terms of an expansion in powers
of  $m/T$ for the case when $T\gg m$. We also point out the importance of
considering that the external frequency takes on discrete values and therefore
implementing the periodicity of the resulting expressions before taking the
analytical continuation to arbitrary complex values. In Sec.~\ref{IV} we
perform the calculation using the MST which involves the use of the Feynman
parametrization. We underline the importance of carrying out the integral over
the Feynman parameter $x\in [0,1]$ to avoid the appearance of spurious endpoint
singularities. We carefully show how the simple Feynman parametrization has to
be corrected to account for the case where the sum of denominators in the
Feynman formula vanishes and emphasize that this correction term accounts for
the whole dependence on the way the momentum approaches zero, in agreement
with the analysis in Ref.~\cite{Weldon}. In Sec.~\ref{V} we apply the results
for the computation of integrals describing the self-energy of a neutral
scalar interacting with charged ones in the presence of a magnetic field. We
finally present our conclusions and give an outlook in Sec.~\ref{concl}. We
leave for the appendices the demonstration of important intermediate
results and alternative derivations of the calculations arising in the
discussion of Secs.~\ref{III} and~\ref{IV}. 

\section{Non-analyticity and mishaps with Feynman parametrization}\label{II} 

The problem of the non-analyticity of thermal field theory calculations as the
momentum components approach zero, has been analyzed by several authors. The
landscape of findings is, at first sight, rather blurred since there are many
details in the calculations that are sources of more extended
discussions. Among these we can mention: the implementation of  
derivative expansion techniques; the validity of perturbative and derivative
expansions exchange and the implementation of the external bosonic field
periodicity. Other studies are concerned with the correct analytic
continuation to arbitrary complex values of the external frequency; the
soundness of some redefinition of variables inside potentially divergent
integrals; the physical interpretation of the imaginary part of the thermal
bubble and the use of the Feynman parametrization.

In this work, our main purpose is to show that the use of Feynman
parametrization, within the MST, is a well defined
procedure. Nevertheless, it is worth pausing to summarize, from a wider
perspective, what has been found in the context of the non-analyticity of
thermal self-energies at the origin in calculations that do not resort to the
use of the MST. 

In general, this problem has been dealt in terms of perturbative and
non-perturbative approaches and both in the ITF and the real time formulation
(RTF) of thermal field theory. The coincidence between perturbative and
non-perturbative calculations is sometimes taken as a guide to decide on the
correctness of the approach. 

The discrepancy between the results in the infrared and the static limits
caused a great deal of confusion, prompting a number of possible
explanations. These ranged from assigning validity only to the infrared 
limit as a genuine result in thermal equilibrium~\cite{Evans}, passing by
suggesting that it is not necessary to assume that the external field is in
thermal equilibrium~\cite{Gribosky, Evans2, Weldon2}, to dismissing the
non-analyticity by claiming that this is 
not present in an exact solution to the slow motion approximation of the
Green's function~\cite{Gribosky, Evans}. In Ref.~\cite{Gribosky}, the
calculation in the ITF is done by extending the external frequency to the
whole imaginary axis and then analytically continuing it to the entire complex
plane. With this procedure, the periodicity of the functions in the external
frequency --that was present before analytical continuation-- is lost. This is
how the erroneous result, that the infrared and static limits coincide, is
obtained. This result seemingly confirmed the one in Ref.~\cite{Bedaque} which
was performed in the RTF. Reference~\cite{Metikas} points out that truncating
the derivative expansion at the beginning of the calculation, either by
keeping only the constant term~\cite{Dolan} or at higher order~\cite{Gribosky,
Evans}, gives misleading results, in the first case, because the operator
nature of the background field is lost; in the latter, because the periodicity
is not considered.

However, Weldon~\cite{Weldon} showed that the results in
Refs.~\cite{Gribosky, Bedaque} go wrong, performing the calculation both in
the ITF and the RTF. In addition to providing physical
arguments for the inequivalence of the infrared and static limits he
demonstrated that in the RTF calculation, the use of the Feynman
parametrization, as is commonly implemented in $T=0$ calculations, needs to be
corrected. The correction accounts for the fact that the real time Feynman
amplitude is not the boundary value of a single analytic function and thus it
is necessary to perform one calculation for the real and another one for the
imaginary part. However, for the discussion concerning the ITF, Weldon argued
that starting from an expression that uses the Feynman parametrization, the
analytical continuation is not unique and leads to a function containing
branch points and an endpoint singularity that need to be removed by the
addition of an extra term. The argument is based on an expression where the
integration interval for the Feynman parameter $x$ has been changed from $x\in
[0,1]$ to $x\in [0,1/2]$ which uses the symmetry of the integrand about
$x=1/2$, {\it before} the analytical continuation is implemented.

In what follows, we will show that the Feynman parametrization can be
implemented in the ITF (within the MST) without introducing spurious branch
points and endpoint singularities. The key ingredients are the implementation
of the periodicity in the expressions {\it before} the analytical
continuation, the use of the original integration interval for the Feynman
parameter $x$ and the accounting of the extra term that corrects the
original Feynman formula when the sum of denominators vanishes and which
happens naturally in the MST. Before proceeding
with this analysis, it is convenient to set the stage and perform an explicit
calculation using the standard technique in the ITF to have a reference to
compare with the result obtained after introducing the MST.

\section{Standard calculation}\label{III}

We start the discussion with the explicit expression for the one-loop
self-energy of a scalar field $\phi$ with a self-interaction of the form
$\lambda\phi^3$. This is given by
\be
   \Pi(p_{0l},p)&=&\frac{\lambda^2}{2} T \sum_{n=-\infty}^{\infty}
   \int \frac{d^3k}{(2\pi)^3} 
   \nonumber\\
   && \Delta (k_{0n},E_k)\Delta (k_{0n}-p_{0l},E_{k-p})
\label{weldon1}
\ee
where $E_x^2={\bf x}^2+m^2$, $p_{0l}=2i\pi l T$, $k_{0n}=2i\pi n T$, with
$l,n$ being integers and
\be
   \Delta (p_{0l},E_p)&=&-\frac{1}{p_{0l}^2-E_p^2}\nonumber\\
   &=&-\sum_{s=\pm 1}\frac{s}{2E_p}\frac{1}{p_{0l}-sE_p}.
   \label{delta}
\ee 
The standard calculation~\cite{Weldon} is done by first summing over the
Matsubara frequencies. This is most easily accomplished by using
Eq.~(\ref{delta}) to write Eq.~(\ref{weldon1}) as 
\be
   \Pi(p_{0l},p)&=&-\frac{\lambda^2}{2} T \sum_{n=-\infty}^{\infty}
       \int \frac{d^3k}{(2\pi)^3} 
       \sum_{r,s=\pm 1} \frac{r s}{4E_kE_{k-p}}\nonumber\\
       &\times&\frac{1}{[2\pi nT+isE_k]
       [2\pi nT+ip_{0l}+irE_{k-p}]}\nonumber\\
\label{weldon1b}
\ee
The sum over the Matsubara frequencies can be computed by means of the
identity 
\be
   \sum_{n=-\infty}^{\infty}\frac{1}{(n+ix)(n+iy)}&=&
   \left(\frac{\pi}{x-y}\right)\nonumber\\
   &\times&
   [\coth(\pi x)-\coth(\pi y)],
\label{weldon1c}
\ee
which yields the expression for the self-energy
\be
   \Pi(p_{0l},p)&=&-\frac{\lambda^2}{2} 
   \int \frac{d^3k}{(2\pi)^3}\nonumber\\ 
   &\times&\sum_{r,s=\pm 1}\left(\frac{1}{8 E_k E_{k-p}}\right)
   \frac{1}{sE_k-(rE_{k-p}+p_{0l})}\nonumber \\
   &\times& 
   \left[ r\coth\left(\frac{E_k}{2T}\right)-
          s\coth\left(\frac{E_{k-p}}{2T}\right)\right],
\label{weldon1e}
\ee
where we used that $\coth(x+il\pi)=\coth(x)$, in account of the fact
that $p_{0l}=2i\pi lT$ and furthermore that $s\coth(sx)=\coth(x)$. We
emphasize that this is an important step necessary to implement the
periodicity in the expression and that eventually allows the analytic
continuation of the result to arbitrary complex values of $p_{0l}$. As
discussed in Ref.~\cite{Metikas} when this condition is not taken, the 
result cannot be interpreted on physical grounds. To stress
the importance of this point, we show in Appendix I that, had
this condition not been taken, the eventual analytical continuation would have
lead to an erroneous result that needs to be corrected precisely by the
addition of the function $\Pi_\delta$ found in Ref.~\cite{Weldon}. Simplifying
Eq.~(\ref{weldon1e}) we 
get 
\be
    \Pi(p_{0l},p)&=&-\frac{\lambda^2}{2} \sum_{s=\pm 1}
    \int \frac{d^3k}{(2\pi)^3}\nonumber\\
    &\times&\Big\{
    \frac{\coth \left(\frac{E_k }{2 T}\right)}
    {4E_k[\left(E_k -sp_{0l}\right)^2-E_{k-p}^2]}\nonumber\\
    &+&(E_k \leftrightarrow E_{k-p})\Big\}.
\label{weldon3}
\ee
Note that upon the change of variable ${\bf k}-{\bf p}\rightarrow {\bf k}$,
the second term in Eq.~(\ref{weldon3}) reduces to the first one and thus the
complete expression for the self-energy is twice the first term in the above
equation. Carrying out the angular integration we get
\be
   \Pi(p_{0l},p)&=& 
   -\frac{\lambda^2}{2(2\pi)^2} 
   \sum_{s=\pm 1} \int_0^\infty kd k
   \frac{\coth\left(\frac{E_k}{2T}\right)}{4pE_k}\nonumber\\
   &\times&\ln\left(\frac{p_{0l}^2-p^2-2sE_kp_{0l}+2kp}
   {p_{0l}^2-p^2-2 sE_kp_{0l}-2kp}\right).
\label{weldon5}
\ee
At this point we take the analytical continuation in $p_{0l}$ from discrete
imaginary values to arbitrary complex ones, $p_{0l}\rightarrow p_0$ and
explore the limiting behavior of Eq.~(\ref{weldon5}) as the momentum 
components of the vector $P^\mu=(p_{0l},{\bf p})$ approach zero. We specialize
to the case where $p_0$ is real. Since the
result depends on the way the limit is taken, we first set $p_0=\alpha p$ 
\be
     \Pi(\alpha p,p)&=& 
     -\frac{\lambda^2}{2(2\pi)^2} 
       \sum_{s=\pm 1} \int_0^\infty k \, d k
    \frac{\coth \left(\frac{E_k }{2 T}\right)} {4E_k p}
   \nonumber\\
   &\times&\ln\left(\frac{\alpha^2p^2-p^2-2 s\alpha E_k  p+2k p}
                   {\alpha^2p^2-p^2-2 s\alpha E_k  p-2k p}\right),
\label{weldon5a}
\ee
and take the limit $p \rightarrow 0$ by expanding the logarithm around $p=0$,
\be
     \Pi(\alpha p,p)&\stackrel{p\rightarrow 0}{=}& 
     -\frac{\lambda^2}{2(2\pi)^2} 
       \sum_{s=\pm 1} \int_0^\infty k \, d k
    \frac{\coth \left(\frac{E_k }{2 T}\right)} {4E_k p}
    \nonumber\\
   &\times&
   \left[
       \frac{k p \left(\alpha ^2 - 1\right)}{k^2 - 
        \alpha ^2 E_k^2} + \log \left(\frac{ 
     s \alpha E_k + k }{s \alpha E_k - k}\right)\right].
\label{weldon5b}
\ee
After carrying out the sum in Eq.~(\ref{weldon5b}) we get in the limit
$p\rightarrow 0$
\be
   \Pi(\alpha p,p)\stackrel{p\rightarrow 0}{=} 
     -\frac{\lambda^2}{2(2\pi)^2} 
      \int_0^\infty d k
      \frac{\coth \left(\frac{E_k}{2 T}\right)}{2E_k}
      \frac{k^2(\alpha^2-1)}{k^2-\alpha^2E_k^2 }.
\label{weldon6}
\ee
By using the identity
\be
    \coth \left(\frac{E_k }{2 T}\right)=(1+2n(E_k )),
\label{weldon7}
\ee
where $n(E_k)$ is the Bose-Einstein distribution, we can separate the
vacuum and thermal contributions of the above equation. Keeping only the
thermal part we obtain
\be
   \Pi^T(\alpha p,p\rightarrow 0)&=& 
     -\frac{\lambda^2}{2(2\pi)^2} 
      \int_0^\infty d k
      \frac{n(E_k)}{E_k}
      \frac{k^2(\alpha ^2-1)}{k^2-\alpha ^2 E_k^2}.\nonumber\\
\label{weldon8}
\ee
We now follow Ref.~\cite{Dolan} to find the explicit expression for $\Pi$
in the infrared limit at high temperature. The result for arbitrary $\alpha$
is given in Appendix II. Setting  $\alpha =0$ in Eq.~(\ref{weldon8}) we get
\be
   \Pi^T(0,p\rightarrow 0)= 
   \frac{\lambda^2}{4(2\pi)^2} 
   \mu^{1-d} \int d^dk
    \frac{n(E_k)}{E_k},
\label{weldon10}
\ee
where in order to write the integral in $d-$dimensions, we have first extended
the integration domain from $[0,\infty]$ to $[-\infty,\infty]$ and thus
multiplied by $1/2$. The extension of the integral in Eq.~(\ref{weldon8}) to
$d$-dimensions represents a way of handling the infinities involved in the
explicit computation and we should keep in mind that in order to make contact
with Eq.~(\ref{weldon8}), the limit $d\rightarrow 1$ will be eventually
taken. The extension to $d-$dimensions also calls for the 
introduction of the mass scale $\mu$. The angular integration in
Eq.~(\ref{weldon10}) can be done straightforward and the result is
\be
   \Pi^T(0,p\rightarrow 0)&=& 
   \frac{\lambda^2}{4(2\pi)^2} 
   \mu^{1-d} \frac{2 \pi^{d/2}}{\Gamma(d/2)}
   \int_0^{\infty} k^{d-1}d k
   \frac{n(E_k)}{E_k}.\nonumber\\
\label{weldon11}
\ee
Using the identity
\be
   \frac{n(E_k)}{E_k}=-\frac{1}{2E_k}
   +\beta\sum_{n=-\infty}^{\infty}\frac{1}{(\beta E_k)^2+(2\pi n)^2},
\label{weldon12}
\ee
where $\beta =1/T$, Eq.~(\ref{weldon11}) can be written as
\be
   \Pi^T(0,p\rightarrow 0)&=& 
   \frac{\lambda^2}{4(2\pi)^2} 
   \mu^{1-d} \frac{\pi^{\frac{d}{2}}}{\Gamma(\frac{d}{2})}
   \int_0^{\infty} k^{d-2}d k^2 \nonumber\\
   &\times&\left(-
   \frac{1}{2E_k}+
   T\sum_{n=-\infty}^{\infty}
   \frac{1}{E_k^2+(2\pi n T)^2}
   \right). \nonumber\\
\label{weldon13}
\ee
Upon the change of variable
\be
   z=\frac{m^2}{k^2+m^2},
\label{weldon14}
\ee
we get
\be
   \Pi^T(0,p\rightarrow 0)&=& 
   \frac{\lambda^2}{4(2\pi)^2} 
   \mu^{1-d} \frac{\pi^{\frac{d}{2}}}{\Gamma(\frac{d}{2})}
   \left\{
   -\frac{1}{2}m^{d-1}\right.\nonumber\\
   &\times&\int_0^1 dz 
   (1-z)^{\frac{d}{2}-1} z^{-\frac{d+1}{2}}  
   \nonumber\\
   &+&T\sum_{n=-\infty}^{\infty}
   (m^2+(2\pi n T)^2)^{\frac{d-2}{2}}\nonumber\\
   &\times& \left. \int_0^1dz 
   (1-z)^{\frac{d}{2}-1} z^{-\frac{d}{2}}
   \right\}.    
\label{weldon16}
\ee
The integrals in the last expression are well known and can be expressed in
terms of ratios of gamma functions $\Gamma$, namely
\be
   \int_0^1 dz (1-z)^\rho z^\gamma =
   \frac{\Gamma(\rho+1)\Gamma(\gamma+1)}{\Gamma(\rho+\gamma+2)}.
\label{weldon17}
\ee
Therefore, Eq.~(\ref{weldon16}) becomes
\be
   \Pi^T(0,p\rightarrow 0)&=& 
   \frac{\lambda^2}{4(2\pi)^2} 
   \mu^{1-d} \frac{\pi^{\frac{d}{2}}}{\Gamma(\frac{d}{2})}\nonumber\\
   &\times&
   \left\{
   -\frac{m^{d-1}}{2}
   \frac{\Gamma(\frac{d}{2})\Gamma(\frac{1-d}{2})}
   {\Gamma(\frac{1}{2})}
   \right. 
   \nonumber \\
   &+&T\sum_{n=-\infty}^{\infty}
   (m^2+(2\pi n T)^2)^{\frac{d-2}{2}}\nonumber\\
   &\times& \left.
   \frac{\Gamma(\frac{d}{2})\Gamma(1-\frac{d}{2})}
   {\Gamma(1)} 
   \right\}.
\label{weldon18}
\ee
Separating the term with $n=0$ in the sum and keeping in mind that the terms in
the sum are even powers of $n$ we get
\be
   \Pi^T(0,p\rightarrow 0)&=& 
   \frac{\lambda^2}{4(2\pi)^2} 
   2T\mu^{1-d} \pi^{\frac{d}{2}}
   \left\{
   -\frac{1}{4T}m^{d-1}
   \frac{\Gamma(\frac{1-d}{2})}
   {\Gamma(\frac{1}{2})}\right.\nonumber\\
   &+&\Gamma\left(\frac{2-d}{2}\right)\frac{m^{d-2}}{2}\nonumber\\
   &+&
   \Gamma\left(\frac{2-d}{2}\right)
   \left.
   \sum_{n=1}^{\infty}
   (m^2+(2\pi n T)^2)^{\frac{d-2}{2}} 
   \right\}.\nonumber\\
\label{weldon19}
\ee
The first and third terms within the curly brackets in the right hand-side of
the above equation have a singularity when 
$d=1$ that should be isolated. The singularity in the first term arises as the
argument of one of the gamma functions vanishes. The singularity in the third
term is less obvious and we concentrate on it. Defining
\be
   S&\equiv&
    \sum_{n=1}^{\infty} \left[
      (m^2+(2\pi n T)^2)^{\frac{d-2}{2}}\right.\nonumber\\
   &-&\left.(2\pi n T)^{d-2}
      +(2\pi n T)^{d-2}\right]  
    \nonumber \\
    &=&
      (2\pi T)^{d-2}\zeta(2-d)\nonumber\\
    &+&\sum_{n=1}^{\infty} \left[
      (m^2+(2\pi n T)^2)^{\frac{d-2}{2}}-(2\pi n T)^{d-2}\right]
    \nonumber \\
    &=& 
      (2\pi T)^{d-2}\zeta(2-d)\nonumber\\
    &+&\sum_{n=1}^{\infty}(2\pi n T)^{d-2} \left[
      \left(\frac{m^2}{(2\pi n T)^2}+1\right)^{\frac{d-2}{2}}-1\right],
\label{weldon20}
\ee
where $\zeta$ is the Riemann zeta function, which has a simple pole at
$d=1$. In the high temperature limit $T\gg m$, we can approximate the above
expression as
\be
    S&\approx& 
      (2\pi T)^{d-2}\zeta(2-d)\nonumber\\
      &+&\sum_{n=1}^{\infty}(2\pi n T)^{d-2} 
      \left[\frac{d-2}{2}\frac{m^2}{(2\pi n T)^2}\right]
      \nonumber\\
      &=&
      (2\pi T)^{d-2}\zeta(2-d)\nonumber\\
      &+&\frac{d-2}{2}(2\pi T)^{d-4}m^2\zeta(4-d).
\label{weldon21}
\ee
Substituting Eq.~(\ref{weldon21}) into Eq.~(\ref{weldon19}) we get
\be
    \Pi^T(0,p\rightarrow 0)&=& 
    \frac{\lambda^2}{4(2\pi)^2} 
    \mu^{1-d} \pi^{\frac{d}{2}}
    \left\{
    -\frac{1}{2}m^{d-1}
    \frac{\Gamma(\frac{1-d}{2})}
    {\Gamma(\frac{1}{2})} \right. 
    \nonumber\\
    &+&\Gamma\left(1-\frac{d}{2}\right)T m^{d-2}
    +2T \ \Gamma\left(1-\frac{d}{2}\right)\nonumber\\ 
    &\times&\left[
    (2\pi T)^{d-2}\zeta(2-d)\right.\nonumber\\
    &+&\left.\left.\frac{d-2}{2}(2\pi T)^{d-4}m^2\zeta(4-d)
    \right]
    \right\}.
\label{weldon22}
\ee
We now set $d=1-2\epsilon$ and make a series for $\epsilon\rightarrow 0$. 
The $\epsilon$-poles cancel and the expression for $\Pi^T$ at high temperature
and in the infrared limit is 
\be
    \Pi^T(0,p\rightarrow 0)&=&  
    \frac{\lambda^2}{4(2\pi)^2} 
    \left\{
    \frac{\pi T}{m}+
    \ln\left(\frac{m}{2T}\right)+
    \gamma_E \right.\nonumber\\
    &-&\left.\frac{m^2 \zeta (3)}{8 \pi^2 T^2}
    \right\},  
\label{weldon24}
\ee
where $\gamma_E$ is Euler's gamma.

\section{Mellin summation technique and Feynman parametrization}\label{IV}

The MST is a useful tool to compute infinite sums~\cite{Kilbas} of the form
encountered in Finite Temperature calculations in the ITF. The technique
resorts to applying a Mellin transform over the discrete frequency to the
expression involving the sum and afterwords applying the inverse transform to
obtain an identity. In this fashion, the calculation of the sum becomes the
easiest part and the problem reduces to computing the Mellin transform and its
inverse of the remaining expression. 
 
We now show how the combined use of the MST and the
Feynman parametrization leads to the same result as the standard
calculation obtained in Eq.~(\ref{weldon24}). For this purpose, we start from
the expression for the self-energy in Eq.~(\ref{weldon1}) separating the sum
over Matsubara frequencies as
\be
   \sum_{n=-\infty}^{\infty}&=&\sum_{n=-\infty}^{-|l|-1}+\sum_{n=|l|+1}^{\infty}
   +\sum_{n=-|l|}^{+|l|}.
   \label{sumseparated}
\ee
This expression has the advantage of separating the sum into pieces where the
frequencies involved have a definite sign from the one where the frequencies
have a mixed sign. The former is suited for the application of the MST since
this last is an integral transform over a continuous variable restricted to
the positive real axis [see Eqs.~(\ref{mellindef})]. Since the combination of
frequencies appears as a square, all that matters is that this has a definite
sign, either positive or negative. For the latter this is not
possible. Nevertheless the calculation can be performed making use of the
Feynman  parametrization. In addition, notice that by transforming the
original discrete frequencies into a continuous variable, there are no
problems associated to the implementation of periodicity conditions. However,
care has to be taken when the sum of denominators of the original Feynman
amplitude vanishes, leading to a correction term, as discussed by Weldon in
Ref.~\cite{Weldon}. In what follows we analyze these contributions separately.

\subsection{Definite sign frequencies}\label{IVa}
 
To begin, let us concentrate on the first two terms arising from the
separation of the sum over Matsubara frequencies in
Eq.~(\ref{sumseparated}) and define 
\be
   \Pi_1(p_{0l},p)&=&\frac{\lambda^2}{2} T \left(
   \sum_{n=-\infty}^{-|l|-1}+\sum_{n=|l|+1}^{\infty}\right)
   \int \frac{d^3k}{(2\pi)^3} 
   \nonumber\\
   && \Delta (k_{0n},E_k)\Delta (k_{0n}-p_{0l},E_{k-p})\nonumber\\
   &=&\frac{\lambda^2}{2} T \left(
   \sum_{n=-\infty}^{-|l|-1}+\sum_{n=|l|+1}^{\infty}\right)
   \int \frac{d^3k}{(2\pi)^3}\nonumber\\
   &&\int_0^1\frac{dx}{[(1-x)D_2+xD_1]^2},
\label{weldon1-III}
\ee
where we have introduced the Feynman parametrization and thus the
integral over the Feynman parameter $x$. In Eq.~(\ref{weldon1-III})
$D_1=\omega_n^2+E_k^2$, $D_2=(\omega_n - \omega_l)^2+E_{k-p}^2$.
$\omega_{n,l}$ are related to $k_{0n}$ and $p_{0l}$ by $k_{0n}=i\omega_n$,
$p_{0l}=i\omega_l$, with $\omega_n=2\pi n T$ and $\omega_l=2\pi l T$. We shift
the three momentum integration variable ${\bf k}\rightarrow {\bf k}-(1-x){\bf
p}$ and, after making the appropriate renaming 
of the summation index, the expression for $\Pi_1$ can be written as 
\be
   \Pi_1(p_{0l},p)&=&\frac{\lambda^2}{2} T \int_0^1dx
   [S_++S_-],
   \label{Ss}
\ee
where $S_{\pm}$ are defined as
\be
   S_\pm&=& \sum_{n=0}^\infty f(\omega_{n\pm}),\nonumber\\
   \label{fs}
\ee
with
\be
   f(y)&=&\mu^{3-d}\nonumber\\
   &\times&\int \frac{d^dk}{(2\pi)^d}\frac{1}
   {[y^2+k^2+x(1-x)(\omega_l^2+p^2)+m^2]^2},
   \nonumber\\
   \label{Ssdefs}
\ee
and
\be
   \omega_{n\pm}^2&=&(2\pi T)^2(n+|l|+1\pm (1-x)l)^2.
   \label{omegas}
\ee
The quantities $\omega_\pm$ will become the variables over which the Mellin
transform is computed, that is to say $\omega_\pm\rightarrow y$ in
Eq.~(\ref{mellindef}). Notice that the expression for $f$ involves an integral
that for later purposes has been extended to $d$-dimensions, and thus the need
to introduce the mass scale $\mu$. In order to make contact with
Eq.~(\ref{weldon1-III}), this time, as opposed to the discussion after
Eq.~(\ref{weldon8}), we will be interested in taking the limit $d\rightarrow
3$.

We perform the sums $S_\pm$ by means of the Mellin summation technique. In
general, the Mellin transform pair $f(y)$, ${\mathcal{M}}[f;s]$ is given by
\be
   \begin{array}{lr}
   {\mathcal{M}}[f;s]=\int_0^\infty y^{s-1}f(y)dy, & 
   \alpha < {\mbox {Re}}(s) < \beta\ \\
   f(y)=\frac{1}{2\pi i}\int_{c-i\infty}^{c+i\infty}y^{-s}{\mathcal{M}}[f;s]ds &
   \alpha < c < \beta,
   \end{array}
   \label{mellindef}
\ee
where $\alpha$ and $\beta$ are determined by the condition that the first of
the integrals in the above equations converges at $y=0$ and $y=\infty$,
respectively. The variable $y$ contains all the dependence on the summation
variable $n$ and is treated as a continuous variable. By expressing $f(y)$ as
an inverse Mellin transform, we can then perform the summation over $n$. The
problem reduces then to finding the Mellin transform and its inverse of the
remaining expression. 

From Eq.~(\ref{Ssdefs}), it is easy to see that $\alpha=0$ and
$\beta=4-d$. In terms of their Mellin transforms, $S_\pm$ can be expressed as
\be
   S_\pm=\frac{1}{2\pi i}\sum_{n=0}^\infty\int_{c-i\infty}^{c+i\infty}
   \frac{1}{\omega_{n\pm}^s}{\mathcal{M}}[f;s]ds.
   \label{sumexpl}
\ee
The sum over $n$ can be explicitly evaluated, yielding
\be
   \sum_{n=0}^\infty\frac{1}{(n+|l|+1\pm (1-x)l)^s}=
   \zeta (s,|l|+1\pm (1-x)l),\nonumber\\
   \label{zetastobe}
\ee
where $\zeta (a,b)$ is the modified Riemann zeta function. To find out the
Mellin transform of $f$, we observe that the integrand of the first of
Eqs.~(\ref{mellindef}), with $f$ given by Eq.~(\ref{Ssdefs}),
can be thought of overall as an integral in $(s+d)$-dimensions of a function of
the square of an $(s+d)$-dimensional vector 
\be
   K^2=\underbrace{\omega_{n\pm}^2}_{s-{\mbox{dim}}}
   +\underbrace{k_{\ }^2}_{d-{\mbox{dim}}}.
   \label{d+s}
\ee
Such integrals are well known~\cite{Peskin} and the result, after
compensating for the volume of the solid angle when extending the integral
from $d$ to $(s+d)$-dimensions, is
\be
   {\mathcal{M}}[f;s]&=&\frac{\mu^{3-d}}{(2\pi T)^s}
   \frac{\Gamma (s/2)}{2(4\pi)^{d/2}}\frac{\Gamma (2-d/2-s/2)}{\Gamma (2)}
   \nonumber\\
   &\times&\frac{1}{[m^2+x(1-x)(\omega_l^2+p^2)]^{2-d/2-s/2}},
   \label{melexp}
\ee
where $\Gamma$ is the gamma function. Combining the results in
Eqs.~(\ref{zetastobe}) and~(\ref{melexp}) the explicit expression for $S_\pm$ is
\be
   S_\pm&=&\mu^{3-d}
   \left(\frac{1}{2\pi i}\right)\left(\frac{1}{2(4\pi)^{d/2}}\right)
   \frac{1}{\Gamma (2)}\nonumber\\
   &\times&[m^2+x(1-x)(\omega_l^2+p^2)]^{d/2-2}\nonumber \\
   &\times&\int_{c-i\infty}^{c+i\infty}ds\ \zeta (s,|l|+1\pm (1-x)l)
   \Gamma (s/2)\nonumber \\
   &\times&\Gamma\left(2-\frac{d+s}{2}\right)
   \left[\frac{m^2+x(1-x)(\omega_l^2+p^2)}{(2\pi T)^2}\right]^{s/2}.
   \nonumber \\
   \label{sumsfin}
\ee
In order to perform the integral over $s$ in Eq.~(\ref{sumsfin}), we notice
that it is necessary to know whether the term
$[m^2+x(1-x)(\omega_l^2+p^2)]/(2\pi T)^2$ is larger or smaller than 
one. For the present purposes where we work in the high temperature limit and
want to explore the analytic properties of $\Pi$ near the origin, we see that
$[m^2+x(1-x)(\omega_l^2+p^2)]/(2\pi T)^2 < 1$. Notice that this assumption
limits the range of values of the external index $l$ to be $l=0,\pm 1$.
Taking $d=3-2\epsilon$, $\epsilon\rightarrow 0^+$, we can choose $c$ such that
$1<c<1+2\epsilon$, in order to both, comply with the upper bound requirement
for the existence of the Mellin transform, Eq.~(\ref{mellindef}), and to avoid
the pole of $\zeta$ at $s=1$. Therefore, the integration contour can be closed
to the right by a half-circle at infinity. The only singularities within the
integration contour are those of $\Gamma [2-(3-2\epsilon + s)/2]$, namely when
$s=1+2\epsilon+2k$, $k=0,1,2,\ldots$ and the integral over $s$ in
Eq.~(\ref{sumsfin}) can be computed by means of the residue theorem, yielding
\be
   S_\pm&=&\lim_{\epsilon\rightarrow 0^+}\mu^{2\epsilon}
   \frac{(4\pi)^{-3/2+\epsilon}}{\Gamma (2)}(2\pi T)^{-(1+2\epsilon)}
   \nonumber \\
   &\times&\sum_{k=0}^\infty \frac{(-1)^k}{k!}\zeta
   (1+2k+2\epsilon,|l|+1\pm(1-x)l)\nonumber \\
   &\times&\Gamma (1/2+k+\epsilon)
   \left[\frac{m^2+x(1-x)(\omega_l^2+p^2)}{(2\pi T)^2}\right]^k.\nonumber \\
   \label{usingres}
\ee
Notice that for $k>0$, $\zeta (1+2k+2\epsilon,|l|+1\pm(1-x)l)$ has no
singularities since $|l|+1\pm(1-x)l\neq 0,-1,-2\ldots$ For $k=0$, the
singularity is regulated by $\epsilon$. Also, the factor
$[m^2+x(1-x)(\omega_l^2+p^2)]^k$ is non-singular. Therefore we can analytically
continue the functions $S_\pm$ from discrete to arbitrary complex values
$i\omega_l=2\pi iTl\rightarrow p_0$. Upon this analytic continuation
\be
   (\omega_l^2+p^2)&\rightarrow& -(p_0^2-p^2)\nonumber\\
   l&\rightarrow&-i\frac{p_0}{2\pi T}\nonumber\\
   |l|&\rightarrow&\frac{|p_0|}{2\pi T}.
   \label{analcont}
\ee
We now explore the limiting behavior of Eqs.~(\ref{usingres}) as the momentum 
components of the vector $P^\mu=(p_0,{\bf p})$ approach zero. Since the
result depends on the way the limit is taken, again we restrict ourselves to
real $p_0$ values, set $p_0=\alpha p$ and take the limit $p \rightarrow 0$ for
the argument of the $\zeta$ function.
\be
   &&\zeta [1+2(k+\epsilon ),1+(|\alpha |\mp i(1-x)\alpha)p/2\pi T]
   \stackrel{p\rightarrow 0}{\rightarrow}
   \nonumber\\
   &&\zeta [1+2(k+\epsilon )]-\nonumber\\
   &&(|\alpha |\mp i(1-x)\alpha)(1+2(k+\epsilon ))
   \zeta [2+2(k+\epsilon )]\left(\frac{p}{2\pi T}\right).
   \nonumber\\
   \label{zetalim}
\ee
The integration over the Feynman parameter involves the computation of the
integral
\be
   Z_\pm&=&\int_0^1dx\Big\{
   \zeta [1+2(k+\epsilon ),1+(|\alpha |\mp i(1-x)\alpha)\frac{p}{2\pi T}]
   \nonumber\\
   &\times&\left[\frac{m^2-x(1-x)(\alpha^2-1)p^2}{(2\pi T)^2}\right]^k\Big\}.
   \label{intoverx}
\ee
As we are interested in evaluating the result near the origin and in the high
temperature limit, we make use of the expansion of $\zeta$ in
Eq.~(\ref{zetalim}) into Eq.~(\ref{intoverx}) and evaluate for the first two
terms in the series expansion of Eq.~(\ref{usingres}), namely
$k=0,1$. Defining
\be
   Z_\pm\equiv\sum_{k=0}^\infty Z_\pm^k ,
   \label{Skdef}
\ee
we get
\be
   Z_\pm^0&=&\zeta [1+2\epsilon]\nonumber\\
   &-&(|\alpha |\mp i\alpha /2)(1+2\epsilon )
   \zeta [2+2\epsilon]\left(\frac{p}{2\pi T}\right)
   \nonumber\\
   &+&{\mathcal{O}}(p^2)\nonumber\\
   Z_\pm^1&=&\zeta [3+2\epsilon ]\left(\frac{m}{2\pi T}\right)^2
   \nonumber\\
   &-&(|\alpha |\mp i\alpha /2)(3+2\epsilon )\zeta [4+2\epsilon ]
   \left(\frac{m}{2\pi T}\right)^2\left(\frac{p}{2\pi T}\right)
   \nonumber\\
   &+&{\mathcal{O}}(p^2).
   \label{intoverxwithk}
\ee  
Notice that in the limit $p\rightarrow 0$, the result coming from the sums
$S_\pm$ is independent of $\alpha$. Combining
Eqs.~(\ref{usingres}),~(\ref{Skdef}) and~(\ref{intoverxwithk}) into
Eq.~(\ref{Ss}), we obtain 
\be
   \Pi_1(0,p\rightarrow 0)&=& 
   \frac{\lambda^2}{4(2\pi)^2}\nonumber\\
   &\times&
   \left[\frac{1}{2\epsilon}+\ln\left(\frac{\mu}{2\sqrt{\pi}T}\right)+
   \frac{\gamma_E}{2}-\frac{m^2\zeta (3)}{8\pi^2T^2} \right].\nonumber\\
\label{pi1fin}
\ee

\subsection{Mixed sign frequencies}\label{IVb}

We now turn to the computation of the third term arising from the separation
of the sum over Matsubara frequencies in Eq.~(\ref{sumseparated}). We define
\be
   \Pi_2(p_{0l},p)&=&\frac{\lambda^2}{2} T 
   \sum_{n=-|l|}^{+|l|}
   \int \frac{d^3k}{(2\pi)^3}\nonumber\\
   &&\int_0^1\frac{dx}{[(\omega_n-x\omega_l)^2+y]^2},
\label{third-fourth}
\ee
where $y=({\bf k}-x{\bf p})^2-x(1-x)(p_{0l}^2-{\bf p}^2)+m^2$.
In order to perform the sum over Matsubara frequencies, it is convenient
to note that the degree in the denominator of Eq.~(\ref{third-fourth}) can be
reduced since
\be
   \frac{1}{[(\omega_n-x\omega_l)^2+y]^2}&=&-\frac{\partial}{\partial m^2}
   \frac{1}{[(\omega_n-x\omega_l)^2+y]}\nonumber\\
   &=&-\frac{\partial}{\partial m^2}
   \sum_{s=\pm 1}\frac{i s}{2y^{1/2}}\nonumber\\
   &\times&\frac{1}
   {[(\omega_n-x\omega_l)+ i s y^{1/2}]},
\label{reduce}
\ee
where in the last line we have resorted to partial fractioning. With this
reduction, the sum over Matsubara frequencies can be easily performed and we
get 
\be
   \Pi_2(p_{0l},p)&=&-\frac{\lambda^2}{2} 
   \int\frac{d^3k}{(2\pi)^3}\int_0^1dx\nonumber\\
   &\times&\frac{\partial}{\partial m^2}
   \sum_{s=\pm 1}\frac{1}{4y^{1/2}}
   \left\{
   \coth\left(\frac{s x p_{0l}}{2T}+\frac{y^{1/2}}{2 T}\right)
   \right.  
   \nonumber\\
   &+&
   \frac{i s}{\pi}
   \psi
   \left(1+|l|+\frac{i s}{\pi}
   \left[\frac{s x p_{0l}}{2T}+\frac{y^{1/2}}{2T}\right]\right)\nonumber\\
   &-&
   \left.
   \frac{i s}{\pi}\psi
   \left(1+|l|-\frac{i s}{\pi}
   \left[\frac{s x p_{0l}}{2T}+\frac{y^{1/2}}{2T}\right]\right)
   \right\},\nonumber\\
\label{almostp}
\ee
where $\psi$ is the digamma function. Before proceeding further, it is
convenient to note that given a function $F$ with argument $(sp_{0l}
x+y(x,m)^{1/2})/2T$, the identity 
\be
   &&\frac{\partial}{\partial m^2}
   \left[\frac{F\left(\frac{s p_{0l} x}{2T}
   +\frac{y(x,m)^{1/2}}{2 T}\right)}{y(x,m)^{1/2}}\right]
   =\nonumber\\
   &&\frac{\partial} {\partial x} 
   \left[\frac{F\left(\frac{s p_{0l} x}{2T}
   +\frac{y(x,m)^{1/2}}{2 T}\right)}{2 y(x,m) 
   \left(s p_{0l}+\frac{1}{2y(x,m)^{1/2}}
   \frac{\partial y(x,m)}{\partial x}\right)}
   \right],
\label{identityF}
\ee
is satisfied. Using Eq.~(\ref{identityF}) into Eq.~(\ref{almostp}), the
integration over the Feynman parameter becomes trivial and we get, after
performing the angular integration
\be
   \Pi_2(p_{0l},p)&=& -\frac{\lambda^2}{2}\int_0^\infty
   \frac{k^2dk}{(2\pi)^2}\sum_{s=-\pm 1}\nonumber\\
   &\times&
   \frac{1}{8E_kpk}\ln\left(\frac{p_{0l}^2-p^2-2sp_{0l}E_k+2kp}
                                      {p_{0l}^2-p^2-2sp_{0l}E_k-2kp}\right)
   \nonumber\\
   &\times& 
   \left[\coth\left(\frac{E_k}{2 T}\right) + 
         \coth\left(\frac{E_k}{2 T}-\frac{s p_{0l}}{2T}\right)\right.
   \nonumber\\
   &+&\frac{i s}{\pi}\psi\left(1+|l|+\frac{i s}{\pi}
   \left[\frac{E_k}{2T}-\frac{s p_{0l}}{2T}\right]\right)
   \nonumber\\
   &-&\frac{i s}{\pi}\psi\left(1+|l|-\frac{i s}{\pi}
   \left[\frac{E_k}{2T}-\frac{s p_{0l}}{2T}\right]\right)
   \nonumber\\           
   &+&\frac{i s}{\pi}\psi
   \left(1+|l|+\frac{i s}{\pi}\frac{E_k}{2T}\right)\nonumber\\
   &-&\left.
   \frac{i s}{\pi}\psi\left(1+|l|-\frac{i s}{\pi}\frac{E_k}{2T}
   \right)\right].
\label{s3e}
\ee
Notice that all along this part of the calculation, we have distinguished
between $l$ and $p_{0l}$ which are in principle related through
$p_{0l}=2i\pi lT$. The reason is that the first term arises as a consequence
of our treating this partial sum over Matsubara frequencies as having $|l|$ in
the limits of the summation index, whereas the second is the discrete
value taken by the external energy. If we now use that $p_{0l}=2i\pi lT$ we
notice in particular that the periodicity in Eq.~(\ref{s3e}),
coming from the argument of the second $\coth$, gets accounted for. However,
there is an extra term that exhibits periodicity in $p_{0l}$ and that is not
that evident from the expression in Eq.~(\ref{s3e}). This comes from the
difference of the functions $\psi$ that have $p_{0l}$ in their argument. This
periodicity can be evidenced by resorting to the identity
\be
   \psi (x+iy) - \psi (x-iy)&=&\sum_{k=0}^\infty\frac{2iy}{y^2+(x+k)^2}
   \nonumber\\
   &=& i\left(\pi\coth[\pi y]-\frac{1}{y}\right)\nonumber\\
   &-&\sum_{k=1}^{|l|}\frac{2iy}{y^2+k^2}.
\label{perioevid}
\ee
After this simplification which allows to account for the periodicity in
$p_{0l}$, the equation can be analytically continued from 
discrete to arbitrary complex values of $p_{0l}\rightarrow p_0$, since
Eq.~(\ref{s3e}) is free from singularities. As before, we explore the limiting
behavior of Eq.~(\ref{s3e}) as the momentum components of the vector
$P^\mu=(p_0,{\bf p})$ approach zero. Again, we restrict ourselves to real
values of $p_0$, set $p_0=\alpha p$ and take the limit $p\rightarrow 0$ and we
get
\be
   \Pi_2(\alpha p,p)&\stackrel{p\rightarrow 0}{=}&
   \frac{\lambda^2}{2(2\pi)^2}T\int_0^\infty dk
   \frac{1}{E_k^2}\nonumber\\
   &=&
   \frac{\lambda^2}{2(2\pi)^2}T\left(\frac{\pi}{2m}\right),
\label{S3infra}
\ee
which is independent of $\alpha$. Upon combining the results of
Eqs.~(\ref{pi1fin}) and~(\ref{S3infra}), we obtain
\be
   \Pi(0,p\rightarrow 0)&=& 
   \frac{\lambda^2}{4(2\pi)^2}\left[\frac{\pi T}{m}
   +\frac{1}{2\epsilon}+
   \ln\left(\frac{\mu}{2\sqrt{\pi}T}\right)\right.\nonumber\\
   &+&
   \left.
   \frac{\gamma_E}{2}-\frac{m^2\zeta (3)}{8\pi^2T^2} \right].
\label{pifin}
\ee

\subsection{$\alpha$ dependence}\label{IVc}

We now proceed to discuss the $\alpha$ dependence of the result since, as we
have seen in Secs.~\ref{IVa} and ~\ref{IVb}, this dependence is absent in the
terms calculated so far. 

As it is shown in Ref.~\cite{Weldon}, the usual Feynman parametrization 
formula at finite temperature has to be corrected when the sum of denominators
can vanish. The correct expression is in this case
\be
   \frac{1}{D_1D_2}&=&\int_0^1\frac{dx}{[(1-x)D_2+xD_1]^2}\nonumber\\
                   &+& 4\pi i \frac{\delta (D_1+D_2)}{D_1-D_2},
\label{correction}
\ee 
where the first term is taken as the principal value. Since within the MST,
the Matsubara frequencies in the sum are first treated as continuous upon the
Mellin transform, and an analytic continuation is required, there is the
possibility that the second term contributes. We proceed to show that this is
indeed the case and that this last term carries the full $\alpha$ dependence
of the result. In Appendix II we show that the term here computed coincides
with the one obtained by using the standard procedure. 

We first look at the contribution from the positive sign frequencies which can
be written as
\be
   \Pi_\alpha^{n>0}&=&\left(\frac{\lambda^2}{2}\right)
   4\pi i T\sum_{n=1}^\infty\int\frac{d^3k}{(2\pi)^3}\nonumber\\
   &\times&\frac{\delta [(\omega_n-\omega_l)^2 + E_{k-p}^2+ \omega_n^2+E_p^2]}
   {(\omega_n-\omega_l)^2 + E_{k-p}^2 - \omega_n^2 - E_p^2}.
\label{positivedel}
\ee
Upon the change of variable
\be
   {\bf k}\rightarrow {\bf k} + {\bf p}/2,
\label{changek}
\ee
and the introduction of the Mellin transform and its inverse,
Eq.~(\ref{positivedel}) can be written as
\be
   \Pi_\alpha^{n>0}&=&\left(\frac{\lambda^2}{2}\right)
   4\pi i T\sum_{n=1}^\infty\int_{c-i\infty}^{c+i\infty}\frac{ds}{\omega_n^s}
   (-i)^s\int_0^\infty du u^{s-1}\nonumber\\
   &\times&\int\frac{d^3k}{(2\pi)^3}
   \frac{\delta [(-iu-\omega_l)^2 - u^2 + 2E_k^2 + p^2/2]}
   {(-iu-\omega_l)^2 + u^2 - 2{\bf p}\cdot{\bf k}},\nonumber\\
\label{Mellinalpha}
\ee
where the Mellin transform has been taken from the discrete Matsubara
frequency $\omega_n\rightarrow -iu$, in anticipation for the taking of the
analytical continuation $i\omega_l\rightarrow p_0$. Also, for convergence of
the integral, $0<c<3$. Taking the analytic continuation $i\omega_l\rightarrow
p_0$, setting $p_0=\alpha p$, the integral over $k$ can be performed
straightforward. In the limit $p\rightarrow 0$ we get
\be
   &&\int\frac{d^3k}{(2\pi)^3}
   \frac{\delta [-(u-\alpha p)^2 - u^2 + 2E_k^2 + p^2/2]}
   {-(u-\alpha p)^2 + u^2 - 2{\bf p}\cdot{\bf k}}
   \nonumber\\
   &=&\frac{1}{(2\pi)^28p}
   \left\{\ln\left[\frac{\sqrt{u^2-m^2}+\alpha u}
   {-\sqrt{u^2-m^2}+\alpha u}\right]\right.\nonumber\\
   &-&\left.\frac{\alpha^2m^2p}
   {\sqrt{u^2-m^2}(m^2-u^2(1-\alpha^2))}\right\}.
\label{intoverk}
\ee
It is easy to show that the potentially dangerous first term in the above
equation for $p\rightarrow 0$ is canceled from a similar 
contribution arising from the sum over negative Matsubara frequencies. We thus
just concentrate in the second term of Eq.~(\ref{intoverk}). We give explicit
results for the case $\alpha > 1$; the case $\alpha <1$ can be worked out by
resorting to the transformation formulas for the hypergeometric
function. Here we jut point out that when $0<\alpha <1$, the high temperature
expansion contains an imaginary part. Integration over the variable $u$ gives
\be
   &&-\int_0^\infty (-i)^sdu u^{s-1}\frac{\alpha^2m^2}
   {\sqrt{u^2-m^2}(m^2-u^2(1-\alpha^2))}\nonumber\\
   &=&(-i)^s\frac{m^{s-1}\sqrt{\pi}}{2}
   \frac{\alpha^2}{(1-\alpha^2)}
   \nonumber\\
   &\times&\left\{
   \frac{\Gamma\left(\frac{3-s}{2}\right)}{\Gamma\left(\frac{4-s}{2}\right)}
   \ _2F_1\left(1,\frac{3-s}{2},\frac{4-s}{2},\frac{1}{1-\alpha^2}\right)
   \right.
   \nonumber\\
   &-&i\frac{\Gamma\left(\frac{s-2}{2}\right)}
   {\Gamma\left(\frac{s-1}{2}\right)}
   \ _2F_1\left(1,\frac{3-s}{2},\frac{4-s}{2},\frac{1}{1-\alpha^2}\right)
   \nonumber\\
   &+&
   i\frac{\Gamma\left(\frac{s}{2}\right)\Gamma\left(\frac{2-s}{2}\right)}
   {\Gamma\left(\frac{1}{2}\right)}\frac{(-1)^{s/2}}{(1-\alpha^2)^{s/2-1}}
   \nonumber\\
   &\times&\left.
   \ _2F_1\left(\frac{s}{2},\frac{1}{2},\frac{s}{2},\frac{1}{1-\alpha^2}\right)
   \right\}
\label{intoveru}
\ee
where $_2F_1$ is a hypergeometric function. The remaining $s$-dependent factor
comes from the summation over the Matsubara frequencies in
Eq.~(\ref{Mellinalpha}) which yields
\be
   \sum_{n=1}^\infty\frac{1}{\omega_n^s}=\frac{1}{(2\pi T)^s}\zeta (s).
\label{sumforZ}
\ee
Therefore, the integral over $s$ involves the terms
\be
   L_1&=&\frac{T}{m}\frac{\alpha^2\sqrt{\pi}}{2(2\pi i)}
   \int_{c-i\infty}^{c+i\infty}ds\left(\frac{-im}{2\pi T}\right)^s
   \frac{\zeta (s)}{(1-\alpha^2)}
   \frac{\Gamma\left(\frac{3-s}{2}\right)}{\Gamma
   \left(\frac{4-s}{2}\right)}
   \nonumber\\
   &\times&\ _2F_1\left(1,\frac{3-s}{2},\frac{4-s}{2},\frac{1}
   {1-\alpha^2}\right)
   \nonumber\\
   L_2&=&i\frac{T}{m}\frac{\alpha^2\sqrt{\pi}}{2(2\pi i)}
   \int_{c-i\infty}^{c+i\infty}ds\left(\frac{-im}{2\pi T}\right)^s
   \frac{(-1)^{s/2}\zeta (s)}{(1-\alpha^2)^{s/2}}
   \nonumber\\
   &\times&\frac{\Gamma\left(\frac{s}{2}\right)
   \Gamma\left(\frac{2-s}{2}\right)}
   {\Gamma\left(\frac{1}{2}\right)}
   \sqrt{\frac{\alpha^2-1}{\alpha^2}}
   \nonumber\\
   L_3&=&i\frac{T}{m}\frac{\alpha^2\sqrt{\pi}}{2(2\pi i)}
   \int_{c-i\infty}^{c+i\infty}ds\left(\frac{-im}{2\pi T}\right)^s
   \frac{\zeta (s)}{(1-\alpha^2)}
   \frac{\Gamma\left(\frac{s-2}{2}\right)}{\Gamma\left(\frac{s-1}{2}\right)}
   \nonumber\\
   &\times&
   \ _2F_1\left(1,\frac{3-s}{2},\frac{4-s}{2},\frac{1}{1-\alpha^2}\right).
\label{threeints}
\ee
Where we used that 
\be
   \ _2F_1\left(\frac{s}{2},\frac{1}{2},\frac{s}{2},\frac{1}{1-\alpha^2}\right)
   =\sqrt{\frac{\alpha^2-1}{\alpha^2}}.
\label{idenF}
\ee

In order to compute $L_{1,2}$ we can close the contour of integration
by a half circle at Re$(s)\rightarrow \infty$ since the convergence of the
integrals is controlled by the ratio $m/T$ which is taken to be less than
one. The integral over this half circle vanishes. $L_{1,2}$ are
given by the residue of the poles of $\Gamma [(3-s)/2]$ and $\Gamma [(2-s)/2$,
which are located at $s=2k+3$, and at $s=2(k+1)$, $k=0,1,2\ldots$,
respectively. Choosing $c>1$ we can avoid the pole of $\zeta (s)$ at
$s=1$. Working with this choice we get. 
\be
   L_1&=&-\frac{i}{2\sqrt{\pi}}\left(\frac{\alpha^2}{\alpha^2-1}\right)
   \sum_{k=0}^\infty\frac{1}{k!}
   \left(\frac{m}{2\pi T}\right)^{2k+2}\nonumber\\
   &\times&\zeta (2k+3)
   \frac{\ _2F_1(1,-k,1/2-k,\frac{1}{1-\alpha^2})}{\Gamma (1/2-k)}.
   \nonumber\\
   L_2&=&\frac{i}{2\pi}\left(\frac{\alpha^2}{\alpha^2-1}\right)^{1/2}
   \sum_{k=0}^\infty\frac{1}{(\alpha^2-1)^{k}}
   \left(\frac{m}{2\pi T}\right)^{2k+1}\nonumber\\
   &\times&\zeta [2(k+1)].
\label{I1I2}
\ee
For $L_3$ we notice that the contour of integration can be closed by a half
cicle at Re$(s)\rightarrow -\infty$ since, as we will show, the contribution
will be proportional to $T/m$ which we take to be larger than one. The
integral over this half circle vanishes. The integral is given by the residue
of the poles of $\Gamma [(s-2)/2]$ which are located at $s=-2k$,
$k=0,1,2\ldots$. However since $\zeta (-2k)$ vanishes for $k=1,2\ldots$, the
only pole that contributes is the one at $s=0$. We thus get
\be
   L_3=i\left(\frac{T}{m}\right)\left(\frac{\alpha}{2}\right)
   \left[\alpha - \sqrt{\alpha^2 - 1}\right].
\label{L3}
\ee

By changing $\omega_n\rightarrow -\omega_n$ in Eq.~(\ref{positivedel}), or
equivalently, $u\rightarrow -u$ in Eqs.~(\ref{Mellinalpha})
and~(\ref{intoverk}), it is straightforward to show that for the modes with
$n<0$ the contribution from the the logarithmic term in Eq.~(\ref{intoverk})
cancels. Therefore, the contribution from the modes with $n\neq 0$ is just
twice the above discussed contribution from the modes with $n>0$. The remaining
term to compute is the one coming from the mode with $n=0$. It is
easy to show that the delta function in this case does not have support and
thus this contribution vanishes.

Writing all together, the final result expressed as an explicit power series
in the ratio $m/2\pi T$ can be written as
\be
   \Pi_\alpha&=&\left(\frac{\lambda^2}{8\pi^2}\right)
   \left\{\left(\frac{\pi T}{m}\right)\alpha
   \left(\sqrt{\alpha^2-1}-\alpha\right)\right.\nonumber\\
   &+&\sqrt{\frac{\alpha^2}{\alpha^2-1}}\left[
   -\left(\frac{m}{2\pi T}\right)\zeta (2)\right.\nonumber\\
   &-&\frac{1}{(\alpha^2-1)}
   \left(\frac{m}{2\pi T}\right)^3\zeta (4)\nonumber\\
   &-&\left.\frac{1}{(\alpha^2-1)^2}
   \left(\frac{m}{2\pi T}\right)^5\zeta (6) - \ldots \right]\nonumber\\
   &+&
   \left(\frac{\alpha^2}{\alpha^2-1}\right)
   \left[\frac{1}{2}\left(\frac{m}{2\pi T}\right)^2\zeta (3)\right.\nonumber\\
   &+&\frac{(3-\alpha^2)}{4(\alpha^2-1)} 
   \left(\frac{m}{2\pi T}\right)^4\zeta (5)\nonumber\\
   &+&\left.\left.\frac{(3\alpha^4-10\alpha^2+15)}{16(\alpha^2-1)^2}
   \left(\frac{m}{2\pi T}\right)^6\zeta (7) + \ldots\right]\right\}.
   \nonumber\\
\label{finalalpha}
\ee
As we show in Appendix II, this result coincides with the one
computed with the standard method.

\subsection{Infrared limit}

As can be seen from Eq.~(\ref{intoverk}) [which is written before computing
the integral over $u$ for which we have assumed $\alpha > 1$] in the infrared
limit ($\alpha=0$), the self energy does not depend on $\alpha$. In order to
compare the result in Eq.~(\ref{pifin}) with Eq.~(\ref{weldon24}), 
which is computed in the infrared limit, we need to subtract from
Eq.~(\ref{pifin}) the vacuum contribution, since the MST does 
not explicitly separate this from the thermal contribution. From
Eqs.~(\ref{weldon6}) and~(\ref{weldon7}) the vacuum contribution to $\Pi$ in
the infrared limit is
\be
   \Pi^{\mbox{vac}}(0,p\rightarrow 0)= 
   \frac{\lambda^2}{8(2\pi)^2}\mu^{1-d}\int\frac{d^dk}{E_k},
\label{vacinfra}
\ee
where we have extended the integral to $d-$dimensions. Notice that in this
case, in order to make contact with Eq.~(\ref{weldon6}), the limit
$d\rightarrow 1$ will be eventually taken. Explicit evaluation of
Eq.~(\ref{vacinfra}) yields
\be
   \Pi^{\mbox{vac}}(0,p\rightarrow 0)=
   \frac{\lambda^2}{8(2\pi)^2} \left\{\frac{1}{\epsilon}
   +\ln\left(\frac{\mu^2}{\pi m^2}\right)-\gamma_E\right\}.
\label{vacinfraexpl}
\ee
Therefore, the thermal contribution is obtained by subtracting
Eq.~(\ref{vacinfraexpl}) from Eq.~(\ref{pifin}) and this is given by
\be
   \Pi^T(0,p\rightarrow 0)&=&\Pi(0,p\rightarrow 0)-
   \Pi^{\mbox{vac}}(0,p\rightarrow 0)\nonumber\\
   &=&
   \frac{\lambda^2}{4(2\pi)^2} \left\{\frac{\pi T}{m}
   +\ln\left(\frac{m}{2T}\right)+\gamma_E\right.\nonumber\\
   &-&\left.\frac{m^2\zeta (3)}{8\pi^2T^2}\right\},
\label{pitermfin}
\ee
which coincides with Eq.~(\ref{weldon24}).

At this point it is important to underline that the ingredient making
possible that the standard procedure reproduced in Sec.~\ref{III} and
the MST described in Sec.~\ref{IV} lead to the same result is the
implementation of the periodicity of the expressions --by appealing to the
fact that the external frequency is discrete-- {\em before} the analytical
continuation to arbitrary complex values of the external frequency is
taken. Also, when the Feynman parametrization is used and afterwards the
periodicity implemented, the 
procedure leads to the well known result, {\it provided the integration
domain for the Feynman parameter $x$ is $x\in [0,1]$.} Nevertheless, as is
discussed in Refs.~\cite{Gribosky, Weldon}, before the analytic continuation,
the integrand is symmetric about $x=1/2$ and thus it is seemingly possible to
get the Feynman integral as twice the result when $x\in[0,1/2]$. We show in
Appendix III that this introduces the extra complication of a spurious
end-point singularity and thus leads to the well known mishaps with the use of
the Feynman parametrization in the ITF.  

We also point out that, as mentioned in Ref.~\cite{Gribosky}, for practical
purposes, the result in Eq.~(\ref{pitermfin}), that is to say, in the infrared
limit, can be directly obtained from Eq.~(\ref{weldon1}) by setting
$p_{0l},p=0$ right from the start. In this case, in the context of the MST,
the mixed frequency sum in Eq.~(\ref{third-fourth}) collapses to the
computation of the contribution of the $n=0$ Matsubara frequency and the
definite sign frequency sums in Eq.~(\ref{weldon1-III}) can be condensed into
a single sum over positive definite frequencies. For calculations involving
propagators raised to higher powers, where one seeks an answer in the infrared
limit, this simplification makes the MST to be a rather convenient
technique, particularly in the high temperature limit $T\gg m$ since it gives
the final answer in terms of a series in $m/T$. We proceed to show that this
is the case when computing the self-energy of a scalar particle interacting
with charged scalar particles in the presence of an external magnetic field.  

\section{Application: Scalar self-energy in a magnetic field}\label{V}

\begin{figure}[t!] 
{\centering
{\epsfig{file=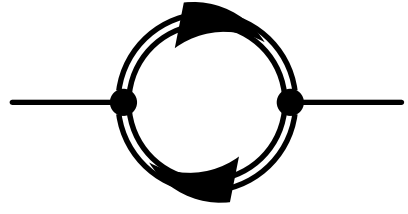, width=0.6\columnwidth}}
\par}
\caption{One-loop Feynman diagram contributing to the self-energy of the
physical Higgs, represented by the single line, interacting with the charged
Higgs components, represented by the double lines, in the presence of an
external weak magnetic field.}
\label{energyzero}
\label{fig1}
\end{figure}

In the Standard Model after symmetry breaking, there is an interaction term
of the physical Higgs $\phi$ with the charged ones $\varphi^\pm$ of the form
$\lambda\varphi^\dagger\varphi\phi$. In the presence of an external
magnetic field, the propagators for the charged modes are affected, becoming,
in the weak field limit and at finite temperature~\cite{nosotros}
\be
   D^B(\omega_n,k)&=&\frac{1}{(\omega_n^2+E_k^2)}\nonumber\\
   &\times&\left(1-
   \frac{(eB)^2}{(\omega_n^2+E_k^2)^2}+
   \frac{2(eB)^2k_\perp^2}{(\omega_n^2+E_k^2)^3}\right),\nonumber\\
   \label{propcharged}
\ee
where $eB$ is the coupling of the charged scalars to the external magnetic
field. One of the diagrams contributing to the physical Higgs self-energy at
one-loop, depicted in Fig.~\ref{fig1}, is given explicitly by
\be
   \Pi^B(\omega_l,p)&=&\lambda^2T\sum_n\int\frac{d^3k}{(2\pi)^3}\nonumber\\
   &\times&D^B(\omega_n,k)D^B(\omega_n-\omega_l,k-p).
   \label{physhiggsself}
\ee
To lowest order in $eB$, this self-energy becomes
\be
   \Pi^B(\omega_l,p)&=&\lambda^2T\sum_n\int\frac{d^3k}{(2\pi)^3}\left\{ I_{11}
   -(eB)^2 \left[I_{31}\right.\right.\nonumber\\
   &+&\left.\left.
   I_{13}-2k_\perp^2I_{41}-2(k-p)_\perp^2I_{14}\right]
   \right\},
   \label{lowestself}
\ee
where we define
\be
   I_{nm}=\frac{1}
   {[\omega_n^2+E_k^2]^n[(\omega_n-\omega_l)^2+E_{k-p}^2]^m}.
   \label{iab}
\ee
When interested in describing the infrared properties of this self-energy, we
look at the infrared limit which, as previously discussed can be obtained in a
straightforward manner by setting $p_{0l},p=0$. In doing so, we get
\be
   \Pi^B(0,p\rightarrow 0)&=&\lambda^2T\sum_n\int\frac{d^3k}{(2\pi)^3}\left\{
   I_{20}-2(eB)^2\right.\nonumber\\
   &\times&\left.
   \left[I_{40}-2k_\perp^2I_{50}\right]
   \right\}.
   \label{lowestself-infra}
\ee
Notice that the functions $I_{n0}$ are all related through
\be
   I_{n0}=\frac{(-1)^{n-1}}{(n-1)!}\frac{\partial^{n-1}}{\partial (m^2)^{n-1}}
   I_{10}.
   \label{relation}
\ee
The MST technique discussed in Sec.~\ref{IV} can be generalized to the
computation of $I_{10}$ and from Eq.~(\ref{relation}) to all the expressions
involved in Eq.~(\ref{lowestself-infra}). The interested reader is referred to
Refs.~\cite{nosotros} for details and the result at high 
temperature and to lowest order in the magnetic field strength is
\be
   \Pi^B(0,p\rightarrow 0)&=&\lambda^2\left\{
   \frac{2}{(4\pi)^2)}\left[\frac{1}{2\epsilon}+\gamma_E+
   \ln\left(\frac{\mu}{4\pi T}\right)\right]\right.\nonumber\\
   &+&\left.\frac{T}{8\pi m} 
   -\frac{(eB)^2}{64}\left(\frac{T}{\pi m^5} + \frac{1}{T^4}
   \frac{\zeta (5)}{16\pi^6} 
   \right)\right\},\nonumber\\
   \label{result}
\ee
where we have not subtracted the vacuum contribution.

We emphasize that the MST is suited to obtain an expression such as
Eq.~(\ref{result}), namely, an expansion at high temperature starting
from the original expression for the self-energy in the ITF. This is so since
the sum over Matsubara frequencies and the integration over the spatial
components of the momentum can be carried out together in a single step, in a
straightforward manner, right from the very beginning.

\section{Summary and Conclusions}\label{concl}

In this work we have shown that the MST is a well defined method to compute
Feynman integrals in the ITF of finite temperature field theory. The MST is
particularly useful to find the explicit result as a series in
a small parameter, for instance, the ratio $m/T$ at high temperature
and in the infrared limit. The method calls for the use of the Feynman
parametrization which in the past has been linked to problems in the ITF when
the analytical continuation from discrete Matsubara frequencies to arbitrary
complex values is performed. We have also shown that these problems are not
endemic to the Feynman parametrization and have traced back their origin to
$(a)$ not implementing the periodicity of the expressions before analytical
continuation and to $(b)$ changing the domain of integration in the Feynman
parameter from $x\in [0,1]$ to $x\in [0,1/2]$ which introduces a spurious
endpoint singularity. We have explicitly shown that when using the MST, and
the calculation is properly carried out, it leads to the same result obtained
by means of the standard technique in the ITF in the infrared limit. In
particular, we have shown the need to take into account the correction term to
the usual Feynman formula, in order to consider the case when the sum of
denominators vanishes, and that this term is the source of the full
$\alpha$ dependence of the result, in agreement with Ref.~\cite{Weldon}. The
usefulness of the method is illustrated by the computation of the one-loop
self energy in the standard model of the physical Higgs field interacting with
the charged components in the presence of a weak external magnetic field, in
the infrared limit.

\section*{Acknowledgments}

The authors acknowledge useful conversations with J. Navarro and the kind
support provided by {\it Programa de intercambio UNAM-UNISON}. Support for
this work has been received in part by PAPIIT-UNAM under grant numbers
IN116008 and IN112308, and CONACyT under grant number 40025-F.

\section*{Appendix I: Evaluation of $\Pi(p_{0l},p)$ without implementing the
          periodicity}\label{appendixI}  

In this appendix, we aim at furthering the argument on the importance of
having implemented the periodicity in the function $\coth$ in
Eq.~(\ref{weldon1e}), to achieve the proper analytic continuation to arbitrary
complex values of $p_{0l}$ in the evaluation of $\Pi(p_{0l},{\mathbf
p})$. We show here that when $p_{0l}$ {\it is not} taken initially as $i$
times a discrete Matsubara frequency, then, when $p_{0l}\rightarrow p_0$,
where $p_0$ is a continuous arbitrary complex number, one is bound to obtain
an spurious term which needs to be canceled precisely by the addition of the
quantity $\pi_\delta$ of Ref.~\cite{Weldon}. 

Without using that $p_{0l}$ takes on discrete integer values, instead of
arriving at that equation we would have  
\be
   \Pi(p_{0l},p)&=&-\frac{\lambda^2}{2}\int \frac{d^3k}{(2\pi)^3} 
   \nonumber \\ 
   &\times & \sum_{r,s=\pm 1} \left(\frac{r s}{8 E_k E_{k-p}}\right) 
   \frac{1}{s E_k-(r E_{k-p}+p_{0l})}\nonumber \\ 
   &\times & \left[ \coth\left(\frac{s E_k}{2 T}\right)-\coth
   \left(\frac{r E_{k-p}+p_{0l}}{2 T}\right) \right],\nonumber\\
\label{weldon26}
\ee
which can be rewritten as
\be
   \Pi(p_{0l},p)&=&-\frac{\lambda^2}{2} \int \frac{d^3k}{(2\pi)^3} 
   \nonumber \\
   &\times & \sum_{r,s=\pm 1} \left(\frac{r s}{8 E_k E_{k-p}}\right) 
   \frac{1}{s E_k-(r E_{k-p}+p_{0l})}\nonumber \\
   &\times& 
   \left[\coth\left(\frac{s E_k}{2 T}\right)-
   \coth\left(\frac{r E_{k-p}}{2 T}\right)\right.\nonumber\\
   &+&
   \left.\frac{\text{csch}^2\left(\frac{r E_{k-p}}{2 T}\right)}
   {\coth\left(\frac{r E_{k-p}}{2 T}\right)
   +\coth\left(\frac{p_{0l}}{2 T}\right)}\right],
\label{weldon27}
\ee
where we used $\coth(a+b)=  \coth (a)-\text{csch}^2(a)(\coth (a)+\coth
(b))^{-1}$ to separate the dependence on $E_{k-p}$ and $p_{0l}$ in the second
hyperbolic function. Note that, compared to what we had in
Eq.~(\ref{weldon1e}), we now have a third term as a result of not fully
exploiting the periodic properties of the functions involved. 

We now concentrate in the last term in Eq.~(\ref{weldon27}) and show that, 
according to Ref.~\cite{Weldon} and in the limit when $p\rightarrow 0$, this
corresponds to minus the function one needs to add to correct the result. Let
us then call $\Pi_X(p_{0l},p)$ the contribution from the
aforementioned term
\be
   \Pi_X(p_{0l},p)&=&-\frac{\lambda^2}{2} 
        \int \frac{d^3k}{(2\pi)^3} 
       \sum_{r,s=\pm 1} \left(\frac{r s}{8 E_k E_{k-p}}\right)
   \nonumber\\
   &\times&
   \frac{1}{s E_k-(r\
     E_{k-p}+p_{0l})} 
      \nonumber \\
     &\times& 
          \left[\frac{\text{csch}^2\left( \frac{r E_{k-p}}{2 T}\right)}
                 {\coth\left(\frac{r E_{k-p}}{2 T}\right)
                          +\coth\left(\frac{p_{0l}}{2 T}\right)}\right],
\label{weldon28}
\ee
where, upon summing over $s$ we have
\be
   \Pi_X(p_{0l},p)&=&-\frac{\lambda^2}{2} 
        \int \frac{d^3k}{(2\pi)^3} 
       \sum_{r=\pm 1} \left( \frac{-r}{4 E_{k-p}}\right)
   \nonumber\\
   &\times&
   \frac{1}{(r E_{k-p}+p_{0l})^2-E_k^2} 
    \nonumber \\
    &\times& 
          \left[\frac{\text{csch}^2\left(\frac{r E_{k-p}}{2 T}\right)}
                 {\coth\left(\frac{r E_{k-p}}{2 T}\right)
                          +\coth\left(\frac{p_{0l}}{2 T}\right)}\right].
\label{weldon29}
\ee
In order to integrate out the angular contribution, we can perform the momentum
shift ${\bf k}-{\bf p} \to {\bf k}$ so that all the angular dependence will be
in the coefficient rather than in the hyperbolic functions. This allows for a
straightforward integration and we arrive at 
\be
   \Pi_X(p_{0l},p)&=&-\frac{\lambda^2}{2} 
        \int_0^{\infty}  \frac{dk}{(2\pi)^2} 
       \sum_{r=\pm 1} \left( \frac{-r k^2}{4 E_{k}}\right)
      \nonumber \\
     &\times& 
          \left[\frac{\text{csch}^2\left( \frac{r E_{k}}{2 T} \right)}
                 {\coth \left( \frac{r E_{k}}{2 T}\right)
                          +\coth \left( \frac{p_{0l}}{2 T} \right)}\right]
  \nonumber \\
  &\times&
  \frac{1}{2kp}
  \ln\left( \frac{p_{0l}^2-p^2+2rE_k p_{0l}-2kp}
            {p_{0l}^2-p^2+2rE_k p_{0l}+2kp} \right).
\nonumber\\
\label{weldon31}
\ee
We now proceed as in the main body of the paper after Eq.~(\ref{weldon5}). We
take the analytical continuation in $p_{0l}$ from discrete 
imaginary values to arbitrary complex ones $p_{0l}\to p_0$. Since the result
depends on how the limit is explored, we first set $p_0 = \alpha p$. To analyze
the behavior near the origin, we expand the function $\coth$ and the 
logarithm around $p=0$ and, up to linear terms, we obtain
\be
   \Pi_X(\alpha p,p)&\stackrel{p \to 0}{=}&-\frac{\lambda^2}{2} 
        \int_0^{\infty} \frac{d{k}}{(2\pi)^2} 
       \sum_{r=\pm 1} \left( \frac{-r  k}{8 E_{k} p} \right)
      \nonumber \\
     &\times& 
          \left[\frac{\alpha p}{2T}~
                \text{csch}^2\left(\frac{r E_{k}}{2 T}\right)
          \right]
   \nonumber\\
   &\times&
   \ln\left(1-\frac{2k}{k+r \alpha E_k} \right),
\label{weldon32}
\ee
which, after summing over $r$ gives 
\be
   \Pi_X(\alpha p,p)&\stackrel{p\to 0}{=}&-\frac{\lambda^2}{2} 
        \int_0^{\infty} \frac{d{k}}{(2\pi)^2} 
          \left[\frac{\alpha k}{16 E_{k} T}
                ~\text{csch}^2\left(\frac{E_{k}}{2 T}\right)\right]
     \nonumber \\
     &\times& 
    \ln\left(\frac{k-\alpha E_k}{k+\alpha E_k} \right)^2.
\label{weldon33}
\ee
Now, just as we did in Sec.~\ref{III}, we are interested in
having an explicit functional dependence on $\alpha$ of $\Pi_X$.
We can then easily extract the thermal contributions thereby 
knowing how $\Pi_X$ modifies $\Pi$, as was discussed in 
Eq.~(\ref{weldon8}). For this purpose, it is convenient 
to note that the term in the square brackets of Eq.(\ref{weldon33})
can be written in terms of a partial derivative ($\partial_k \text{coth} E_k = 
k E_k^{-1} \text{csch}^2 E_k$), so that we can complete a total derivative
through integration by parts, to have
\be
   \Pi_X(\alpha p,p)&\stackrel{p \to 0}{=}&-\frac{\lambda^2}{2} 
        \int_0^{\infty}  \frac{d{k}}{(2\pi)^2} \left\{
         -\frac{\alpha}{8}\right.
           \nonumber \\
     &\times&
    \frac{\partial}{\partial k}\left[\coth\left(\frac{E_{k}}{2 T}\right)
     \ln\left(\frac{k-\alpha E_k}{k+\alpha E_k} \right)^2
     \right]
   \nonumber \\
   &+&\left.
     \frac{\alpha^2(E_k^2-k^2)}{2E_k(k^2-\alpha^2 E_k^2)}
\coth\left(\frac{E_{k}}{2 T}\right) \right\}.
\label{weldon35}
\ee
Finally we can separate the vacuum and the thermal contributions using
the identity in Eq.~(\ref{weldon7}), so that the thermal part is
\be
   \Pi_X^T(\alpha p,p)&\stackrel{p\to 0}{=}&
   -\frac{\lambda^2}{2 (2\pi)^2} \int_0^{\infty} d{k}
   \frac{n(E_{k})}{E_k} \frac{\alpha^2 m^2}{(k^2-\alpha^2 E_k^2)}.
   \nonumber\\
\label{weldon36}
\eea
The function $\Pi_X^T$ in Eq.(\ref{weldon36}) is precisely 
$\lim_{p \to 0} \pi_\delta(\alpha p,p)$ found in Eq.(30) of
Ref.~\cite{Weldon}, {\it but with the opposite sign}. We can see that in the
event of not implementing the periodicity, as we have analyzed
in this appendix, inevitably we will end up with a contribution stemming from
the extra term $\Pi_X$. The situation is corrected, as noted in
Ref.~\cite{Weldon}, if one adds a function that behaves just as $\Pi_\delta$
in the limit considered. This turns out to be an important observation, since
we are presenting evidence that neglecting the implementation of the
periodicity in the external frequency is linked to the need of such correcting
function. Further developments on this argument are presented in the rest of
this work.

\section*{Appendix II: Evaluation of $\Pi^T(\alpha p,p)$ for 
          $p\rightarrow 0$ and arbitrary $\alpha$ }\label{appendixII}

We start from Eq.(\ref{weldon8}) rewriting it as
\be
   \Pi^T(\alpha p,p)= 
     \frac{\lambda^2}{2(2\pi)^2} 
      \int_0^\infty k^2 d k
      \frac{n(E_k)}{E_k}
      \frac{1}{E_k^2-\alpha'^2m^2},
\label{arbi1}
\ee
where $\alpha'^2=\frac{1}{1-\alpha^2}$. We follow again Ref.\cite{Dolan} and
use the identity in Eq.~(\ref{weldon12}) into Eq.~(\ref{arbi1}), that is
\be
   \Pi^T(\alpha p,p)&=& 
     \frac{\lambda^2}{2(2\pi)^2}\frac{\mu^{3-d}}{4\pi} 
      \int d^d k \frac{1}{E_k^2-\alpha'^2m^2}
     \left\{
      -\frac{1}{2E_k}\right.\nonumber\\
      &+&\left.T\sum_{n=-\infty}^{\infty}\frac{1}{(E_k)^2+(2\pi n T)^2}
      \right\}.
      \label{arbi2}
\ee
where we have written the integral in $d$-dimensions. The first structure in
Eq.~(\ref{arbi2}) is
\be
    J_1= 
     - \mu^{3-d} \frac{\lambda^2}{2(2\pi)^2}\frac{1}{4\pi} 
      \int d^d k
      \frac{1}{2E_k}
      \frac{1}{E_k^2-\alpha'^2m^2}.
\label{arbi3}
\ee
Carrying our the angular integration and upon the change of variable
$z=\frac{m^2}{k^2+m^2}$, we get
\be
    J_1&=& 
     - \mu^{3-d} m^{d-3}\frac{\lambda^2}{2(2\pi)^2}\frac{1}{8\pi} 
      \frac{\pi^{d/2}}{\Gamma(\frac{d}{2})}
      \int_0^1 d z (1-z)^{\frac{d}{2}-1}\nonumber\\
   &\times&
    z^{\frac{1-d}{2}}
      (1-\alpha'^2 z)^{-1}.
\label{arbi4}
\ee
Using the identity
\be
   _2F_1(a,b,c;z)&=&\frac{\Gamma(c)}{\Gamma(b)\Gamma(c-b)}
   \int_0^1dt (1-t)^{c-b-1}\nonumber\\
   &\times&
   t^{b-1}(1-z t)^{-a},
\label{arbi5}
\ee
where $_2F_1$ is the hypergeometric function, we get
\be
   J_1&=& 
     - \mu^{3-d} m^{d-3}\frac{\lambda^2}{2(2\pi)^2}
      \frac{\pi^{d/2}}{8\pi} 
      \frac{\Gamma(\frac{3-d}{2})}{\Gamma(\frac{3}{2})}
      \nonumber\\
      &\times&
      {_2F_1}\left(1,\frac{3-d}{2},\frac{3}{2};\alpha'^2\right).
\label{arbi6}
\ee
For the second structure in Eq.~(\ref{arbi2}), a similar proceedure leads to
\be
   J_2&=&  
     \mu^{3-d}\frac{\lambda^2}{2(2\pi)^2}\frac{\pi^{d/2}}{4\pi}
      \Gamma\left(2-\frac{d}{2}\right)\nonumber\\
      &\times&
      T\sum_{n=-\infty}^{\infty}
      (m^2+\omega_n^2)^{\frac{d}{2}-2}\nonumber\\
      &\times&
      {_2F_1}\left(1,2-\frac{d}{2},2;\frac{\omega_n^2+\alpha'^2 m^2}
      {\omega_n^2+ m^2}\right).
\label{arbi8}
\ee
Note that for the term $n=0$ in Eq.~(\ref{arbi8}), the argument of the
hypergeometric function becomes independent of $m$ and $T$. The
result for $\Pi (\alpha p,p)$ is thus
\be
   \Pi^T(\alpha p,p)= J_1+J_2.
\label{arbi9}
\ee
In order to veryfy this result in the limit $\alpha =0$ ($\alpha'=1$) we
recall the identity
\be
   _2F_1(a,b,c;1)=\frac{\Gamma(c)\Gamma(c-b-a)}{\Gamma(c-b)\Gamma(c-a)},
\label{arbi12}
\ee
that can be used to write
\be
    J_1&\stackrel{\alpha=0}{\rightarrow}&
    -\frac{\lambda^2}{2(2\pi)^2}
      \frac{1}{8\pi} \left(\frac{\mu}{m}\right)^{3-d} 
      \frac{\pi^{(d-1)/2}\Gamma\left(\frac{3-d}{2}\right)
      \Gamma\left(\frac{d}{2}-1\right)}{\Gamma(\frac{d}{2})}
      \nonumber\\
    J_2&\stackrel{\alpha=0}{\rightarrow}&
    \mu^{3-d}\frac{\lambda^2}{2(2\pi)^2}\frac{1}{4\pi}
      \frac{\pi^{d/2}\Gamma\left(2-\frac{d}{2}\right) 
      \Gamma\left(\frac{d}{2}-1\right)}{\Gamma(\frac{d}{2})}
      \nonumber\\
      &\times&T\sum_{n=-\infty}^{\infty}
      (m^2+\omega_n^2)^{\frac{d}{2}-2}.
\label{arbi13}
\ee
Using the procedure as in Eqs.~(\ref{weldon18})-(\ref{weldon21}) to obtain the
high temperature limit, we get for $J_2$
\be
   J_2&=&\frac{\lambda^2}{2(2\pi)^2}\frac{1}{4\pi}\mu^{3-d}
      \frac{\pi^{d/2}\Gamma(2-\frac{d}{2}) 
      \Gamma(\frac{d}{2}-1)}{\Gamma(\frac{d}{2})}
      \nonumber \\
      &\times&\left(\frac{}{}
      T m^{d-4}+2T(2\pi T)^{d-4}\zeta(4-d)\right.\nonumber \\
      &\times&\left.
      +2T\left(\frac{d}{2}-2\right)(2\pi T)^{d-6}m^2\zeta(6-d).
      \right)
\label{arbi15}
\ee
Taking $d\rightarrow 3-2\epsilon$ and $\alpha =0$, the result in the infrared
limit is
\be
    \Pi^T(0,p\rightarrow 0)&=&  
    \frac{\lambda^2}{4(2\pi)^2} 
    \left\{
    \frac{\pi T}{m}+
    \ln\left(\frac{m}{2T}\right)+
    \gamma_E \right.\nonumber\\
    &-&\left.\frac{m^2 \zeta (3)}{8 \pi^2 T^2}
    \right\},
\label{arbi16}
\ee
which coincides with Eq.~(\ref{weldon24}).

We can also use the former analysis to give an explicit expression for the
$\alpha$ dependence of the self-energy in the high temperature limit. We
first separate from Eq.~(\ref{arbi1}) all $\alpha$ dependence. In terms of the
parameter $\alpha'$, we get
\be
     \Pi(\alpha p,p)&=& 
      \frac{\lambda^2}{2(2\pi)^2} 
      \int_0^\infty d k
      \frac{n(\omega_k)}{\omega_k}
      \left[1+
      \frac{(\alpha'^2-1)m^2}{\omega_k^2-\alpha'^2m^2}
      \right]\nonumber\\
      &\equiv& \Pi_0+\Pi_{\alpha},
\label{alphaexp1} 
\ee
where 
\be
   \Pi(\alpha p,p)_\alpha&\equiv&
      \frac{\lambda^2 (\alpha'^2-1)m^2}{2(2\pi)^2}
      \int_0^\infty d k
      \frac{n(\omega_k)}{\omega_k}
      \frac{1}{\omega_k^2-\alpha'^2m^2}.\nonumber\\
\label{alphaexp2}
\ee
Notice that the above integral can be obtained from  Eq.~(\ref{arbi1})
taking $d=1-2\epsilon$ and with the changes
\be
   \mu^{3-d}&\rightarrow& \mu^{1-d} \nonumber \\
   \frac{1}{4\pi}&\rightarrow& \frac{1}{2}
\label{alphaexp3}
\ee
from where we get
\be
   J_1&\rightarrow&
     - \frac{\lambda^2(\alpha'^2-1)}{2(2\pi)^2}
      \frac{1}{4} 
      \frac{\Gamma(\frac{1}{2})}{\Gamma(\frac{3}{2})} 
      {_2F_1}(1,1,\frac{3}{2};\alpha'^2) \nonumber \\
      &=&
     - \frac{\lambda^2(\alpha'^2-1)}{2(2\pi)^2}
      \frac{1}{2} 
      \frac{\sin ^{-1}(\alpha')}{\alpha'\sqrt{1-\alpha'^2}}\nonumber\\
    J_2&\rightarrow&+\frac{\lambda^2(\alpha'^2-1)}{2(2\pi)^2}
      \frac{\pi Tm^{-1}}{2}
      \left[
      {_2F_1}(1,\frac{3}{2},2;\alpha'^2)\right.\nonumber \\
      &+&\left.2\sum_{n=1}^{\infty}
      x_n^{3}(x_n^2+1)^{-\frac{3}{2}}
      {_2F_1}(1,\frac{3}{2},2;\frac{1+\alpha'^2 x_n^2}
                                     {1+ x_n^2})
       \right]    \nonumber    \\
      \Pi(\alpha p,p)_{\alpha}&\equiv& J_1+J_2.
\label{alphaexp4}
\ee
where $x_n=m/2\pi nT$. Notice that $J_1$ in Eq.~(\ref{alphaexp4}) yields a
$T$-independent term and therefore contributes only to the vacuum part. This
can be shown to correspond to considering the pole of $\zeta (s)$ at $s=1$ in
$L_2$ given in Eq.~(\ref{threeints}). We can thus ignore this term. 

In the high temperature limit the parameter $x_n\ll 1$, thus we can perform a
series expansion in $J_2$, yielding
\be
   J_2&=&\left(\frac{\lambda^2}{8\pi^2}\right)
   \left\{\left(\frac{\pi T}{m}\right)\alpha
   \left(\sqrt{\alpha^2-1}-\alpha\right)\right.\nonumber\\
   &+&\sqrt{\frac{\alpha^2}{\alpha^2-1}}\left[
   -\left(\frac{m}{2\pi T}\right)\zeta (2)\right.\nonumber\\
   &-&\frac{1}{(\alpha^2-1)}
   \left(\frac{m}{2\pi T}\right)^3\zeta (4)\nonumber\\
   &-&\left.\frac{1}{(\alpha^2-1)^2}
   \left(\frac{m}{2\pi T}\right)^5\zeta (6) - \ldots \right]\nonumber\\
   &+&
   \left(\frac{\alpha^2}{\alpha^2-1}\right)
   \left[\frac{1}{2}\left(\frac{m}{2\pi T}\right)^2\zeta (3)\right.\nonumber\\
   &+&\frac{(3-\alpha^2)}{4(\alpha^2-1)} 
   \left(\frac{m}{2\pi T}\right)^4\zeta (5)\nonumber\\
   &+&\left.\left.\frac{(3\alpha^4-10\alpha^2+15)}{16(\alpha^2-1)^2}
   \left(\frac{m}{2\pi T}\right)^6\zeta (7) + \ldots\right]\right\}.
   \nonumber\\
\label{finalalphaJ2}
\ee
which coincides with Eq.~(\ref{finalalpha}).

\section*{Appendix III: $\Pi_\delta$ in the ITF}\label{appendixIII}

In this appendix we show that the function $\Pi_\delta$ found in
Ref.~\cite{Weldon} emerges in the ITF making use of the Feynman
parametrization only when the limits of integration are replaced from $x\in
[0,1]$ to $x\in [0,1/2]$. 

We start from Eq.~(3.33) in Ref.~\cite{Gribosky}
\be
   \Pi_x(p_{0l},p)&=&-\frac{\lambda^2}{8}\int\frac{d^3k}{(2\pi)^3}
          \int_0^1dx \nonumber\\
         &\times&
         \frac{\partial}{\partial m^2}
         \sum_{r=\pm 1}
         \frac{\coth \frac{\beta}{2}(r \ xp_{0l}+y^{\frac{1}{2}})}
              {y^{\frac{1}{2}}},
\label{pix}
\ee
where $y$ is defined as
\be
  y=E_k^2+x(E_{k-p}^2-E_k^2)- x(1-x)p_{0l}^2.
\label{cambiodevar}
\ee
It is worth noticing that in Ref.~\cite{Weldon} the change of variable 
${\bf k}-x{\bf p}\rightarrow {\bf k}$ is performed in Eq.~(\ref{pix}), but this
change  is not allowed in this case since the integral is divergent, unless
the divergence is regulated by using for instance, dimensional regularization.  

Using the identity in Eq.~(\ref{identityF}) into Eq.~(\ref{pix}), we get
\be
   \Pi_x(p_{0l},p)&=&-\frac{\lambda^2}{8}
          \sum_{r=\pm 1}
          \int\frac{d^3k}{(2\pi)^3}
          \int_0^1 dx\nonumber\\
          &\times& 
          \frac{\partial}{\partial x}
          \left[
          \frac{\coth \frac{\beta}{2}(r \ xp_{0l}+y^{\frac{1}{2}})}
              {2y (r \ p_{0l}+\frac{\partial y^{\frac{1}{2}}}{\partial x})} 
          \right].
\label{pix2}
\ee
The integral over $x$ becomes trivial and when evaluating in the integration
limits $x=1$, $x=0$ we obtain Eq.~(\ref{weldon1e}). This is what is done in
Ref.~\cite{Gribosky} which leads to the correct result, provided the
periodicity in $\coth$ is imposed, as discussed in Sec.~\ref{III}. However, if
we instead follow Ref.~\cite{Weldon} and use that the integrand is symmetric
about $x=1/2$ and thus that the integral over $x$ in the interval $x\in [0,1]$
is twice the integral in the interval $x\in [0,1/2]$, we get
\be
      \Pi_x(p_{0l},p)&=&-\frac{\lambda^2}{4}
          \sum_{r=\pm 1}
          \int\frac{d^3k}{(2\pi)^3}
          \int_0^\frac{1}{2} dx\nonumber\\
          &\times& 
          \frac{\partial}{\partial x}
          \left[
          \frac{\coth \frac{\beta}{2}(r \ xp_{0l}+y^{\frac{1}{2}})}
              {2y (r \ p_{0l}+\frac{\partial y^{\frac{1}{2}}}{\partial x})} 
          \right].
\label{pix3}
\ee
Notice that Eq.~(\ref{pix3}) is valid when $p_{0l}$ is imaginary and discrete,
since only in this case, $\coth$ is periodic. Evaluating the integral over $x$
in Eq.~(\ref{pix3}) we get
\be
      \Pi_x(p_{0l},p)&=&-\frac{\lambda^2}{4}
          \sum_{r=\pm 1}
          \int\frac{d^3k}{(2\pi)^3}\nonumber\\
          &\times&\left[-
          \frac{\coth \frac{\beta}{2}E_k}
              {E_k(2r\ p_{0l}E_k+E_{k-p}^2-E_k^2-p_{0l}^2)} 
          \right.\nonumber\\
          &+&\left.\left.\frac{\coth \frac{\beta}{2}(r \ xp_{0l}+y^{\frac{1}{2}})}
              {2y (r \ p_{0l}+\frac{\partial y^{\frac{1}{2}}}{\partial x})} 
          \right|_{x=1/2}\right],
\label{pix4}
\ee
where the first term results from evaluating in the lower limit of the
$x$-integral and in the second one we have left indicated that $x$ is
evaluated in $1/2$. Notice that when completing the square in the denominator
of the first term in Eq.~(\ref{pix4}), this becomes identical to the result in
Eq.~(\ref{weldon3}), which is the correct result, thus leaving
Eq.~(\ref{pix4}) with an extra term, which in fact, as we proceed to show,
corresponds to the function $-\Pi_\delta$ in Ref.~\cite{Weldon}. To show this we
must carry out the angular integration in Eq.~(\ref{pix4}). Defining
\be
      \Pi_{\Delta}(p_{0l},p)&\equiv&-\frac{\lambda^2}{4}
          \sum_{r=\pm 1}
          \int\frac{d^3k}{(2\pi)^3}\nonumber\\
          &\times&
          \frac{\coth \frac{\beta}{2}(r \ xp_{0l}+y^{\frac{1}{2}})}
              {y^{\frac{1}{2}}[2r y^{\frac{1}{2}}p_{0l} + 
             E_{k-p}^2-E_k^2- (1-2x)p_{0l}^2]},\nonumber\\
\label{pix5}
\ee
where $x$ should be evaluated in $1/2$. Upon the change of variable ${\bf
k}-x{\bf p} \rightarrow  {\bf k}$, the dependence of the angle inside the
function $\coth$ is removed and we get 
\be
      \Pi_{\Delta}(p_{0l},p)&=&-\frac{\lambda^2}{4}
          \sum_{r=\pm 1}
          \int\frac{d^3k}{(2\pi)^3}\nonumber\\
          &\times&
          \left[
          \frac{\coth \frac{\beta}{2}(r \ xp_{0l}+\phi^{\frac{1}{2}})}
          {\phi^\frac{1}{2}[2r \phi^{\frac{1}{2}}p_{0l}-2 {\bf k} \cdot {\bf p} ]}
          \right],
\label{pix6}
\ee
where $\phi= k^2 +m^2-x(1-x)(p_{0l}^2-p^2)$. The remaining angular integration
is readily performed and the result is 
\be
      \Pi_{\Delta}(p_{0l},p)&=&-\frac{\lambda^2}{4(2\pi)^2}
           \int_0^\infty \frac{k d k}{\phi^\frac{1}{2} p}
            \ln\left(\frac{p_{0l}\phi^{\frac{1}{2}}+kp}{p_{0l}
            \phi^{\frac{1}{2}}-kp}\right)\nonumber\\
       &\times&\left.
       \left[n(xp_{0l}+\phi^{\frac{1}{2}})-
       n(-xp_{0l}+\phi^{\frac{1}{2}})\right]\right|_{x=1/2},\nonumber\\
\label{pix7}
\ee
where $n$ is the Bose-Einstein distribution. Notice that if in
Eq.~(\ref{pix7}) we use that $p_{0l}$ is purely imaginary and discrete, the
function $\Pi_\Delta$ vanishes. However, if $p_{0l}$ is analytically continued
to arbitrary complex values, the correct result is obtained only by the
addition of the function $\Pi_\delta$ found in Ref.~\cite{Weldon}, which
exactly cancels $\Pi_\Delta$.

\end{document}